# Accurate measurement of atomic magnetic moments by minimizing the tip magnetic field in STM-based electron paramagnetic resonance

Tom S. Seifert, Stepan Kovarik, Pietro Gambardella, Sebastian Stepanow

[1] Department of Materials, ETH Zürich, 8093 Zürich, Switzerland

*Corresponding author. Email: tom.seifert@fu-berlin.de; sebastian.stepanow@mat.ethz.ch

Electron paramagnetic resonance (EPR) performed with a scanning tunneling microscope (STM) allows for probing the spin excitation of single atomic species with MHz energy resolution. One of the basic applications of conventional EPR is the precise determination of magnetic moments. However, in an STM, the local magnetic fields of the spin-polarized tip can introduce systematic errors in the measurement of the magnetic moments by EPR. We propose to solve this issue by finding tip-sample distances at which the EPR resonance shift caused by the magnetic field of the tip is minimized. To this end, we measure the dependence of the resonance field on the tip-sample distance at different radiofrequencies and identify specific distances for which the true magnetic moment is found. Additionally, we show that the tip's influence can be averaged out by using magnetically bistable tips, which provide a complementary method to accurately measure the magnetic moment of surface atoms using EPR-STM.

## INTRODUCTION

Electron paramagnetic resonance (EPR) [1] allows for the precise determination of the electronic and magnetic properties of paramagnetic species by measuring their magnetic moment with high accuracy. In typical experiments performed in resonant cavities, the magnetic moment can be measured with an accuracy of up to a few parts per million [2], often limited by the accuracy of measuring the magnetic field or the applicable radiofrequency [3]. Examples include the analysis of the structure-reactivity relationship in heterogeneous catalysis [4, 5], e.g., the sensitive probing of oxygen radicals on the surface of MgO [6], and spin-levels of molecular nanomagnets [7, 8]. The recent implementation of EPR into a scanning tunneling microscope (STM) enabled the study of magnetic properties and interactions of single surface atoms with MHz resolution, which is orders of magnitudes below the thermal limit in typical low-temperature STMs [9-12].

In EPR-STM, a radiofrequency electric field resonantly excites the Zeeman-split magnetic states of a single atom under the STM tip. The resulting change in the occupation of the magnetic states is probed by the spin-polarized STM tip through the tunnel-magneto-resistive effect. The combined spatial resolution and spectroscopic power of EPR-STM opens new avenues for studying surface magnetism and model quantum spin systems [9, 13-15] as well as reactive chemistry at surfaces [3]. However, precise measurements of the magnetic moment of surface atoms remain challenging. On the one hand, recent EPR-STM studies [10, 11, 16-19] reported statistical errors down to $10^{-3}$ for measurements of the magnetic moments of Fe and Ti atoms on Mg(100). On the other hand, measurements of the Fe magnetic moment differ by up to 1 $\mu_B$ in different experiments [16, 17, 19, 20], pointing towards a source of significant systematic errors. A major cause for this inconsistency is the poorly known magnetic field of the spin-polarized STM tip, which adds to the external magnetic field at the position of the adatom (Fig. 1a) [21]. This magnetic field is the cause of systematic errors when the magnetic moment is derived from fits of the Zeeman splitting as a function of external field. Efforts to account for the influence of the tip magnetic field included the extrapolation to infinite tip-sample distance [22], the interpolation between pairs of resonance fields and radiofrequencies [9], the use of spectator atoms [16, 17] or the characterization of the tip magnetic field [23]. All these approaches, however, suffer from specific drawbacks that we will outline further below.

In this work, we demonstrate that the influence of the tip magnetic field on the position of the EPR resonance is minimized at specific tip-sample distances. This occurs when the dipolar field and the exchange interaction of the magnetic tip cancel each other exactly along the diagonal axis of the spin [23]. In the following we refer to this special measurement conditions as no-tip influence (NOTIN) EPR. To find the special NOTIN points experimentally, we compare the resonance field vs tip-sample distance curves acquired at different radiofrequencies. At the intersection points of the curves, i.e., the NOTIN distances, the magnetic moment inferred for the surface atoms is independent of the radiofrequency and thus corresponds to the true magnetic moment.

We demonstrate our approach using the two well-characterized magnetic atoms, namely Fe adsorbed on an oxygen site and hydrogenated Ti (TiH) adsorbed on a bridge site between two oxygen atoms on the MgO/Ag(100) surface (Fig. 1) [10]. The magnetic moments along the out-of-plane direction are found to be $\mu_{Fe} = 5.47 \pm 0.06\ \mu_B$ and $\mu_{TiH} = 0.99 \pm 0.01\ \mu_B$, respectively. Further, we discuss the assumptions and limitations of our approach and benchmark the NOTIN procedure against previously reported measurements of the magnetic moments. Finally, we explore the use of magnetically bistable tips [13, 24] giving rise to the simultaneous observation of two EPR resonances. Averaging the two associated apparent magnetic moments allows us to remove the influence of the magnetic tip. The magnetic moments obtained using bistable tips show excellent agreement with the moments obtained through the NOTIN procedure.

**THEORY**

The basic EPR equation describing the resonance condition reads [1]

$$2\mu B = hf \qquad (1)$$

where $\mu$ is the component of the magnetic moment along the magnetic field **B** at the position of the surface atom, $h$ is Planck's constant and $f$ is the frequency of the applied radiofrequency field. Thus, knowing the pair $B$ and $f$ at resonance the moment $\mu$ can be calculated. In EPR-STM an external magnetic field $\mathbf{B}_{ext}$ is applied to split the energy levels of the atom (Fig. . 1) and a resonant radiofrequency electric field applied to the STM junction induces transitions between these Zeeman-split states [10]. Microscopically, different EPR-STM driving mechanisms have been suggested [9, 23, 25, 26]. Yet, recent experiments strongly suggest that the driving radiofrequency electric field couples piezoelectrically to the adatom leading to a motional oscillation along the surface normal [23]. Due to the inhomogeneous magnetic field of the nearby magnetic STM tip, the surface atom experiences a time-dependent magnetic field, which ultimately drives the EPR transition [23, 25, 27]. The experimental observation of the EPR resonance relies on the tunnel-magneto-resistance effect of the magnetic STM tip, which is sensitive to changes in the magnetic state of the surface atom [28]. Hence, there is an intimate coupling between driving and sensing of the magnetic resonance of the atom. Typical EPR-STM measurements only yield an apparent magnetic moment $\tilde{\mu}$ given by

$$\tilde{\mu} = \frac{hf}{2B_{ext}^0}, \tag{2}$$

where $B_{ext}^0$ is the value of the external magnetic field at resonance measured at constant radiofrequency.

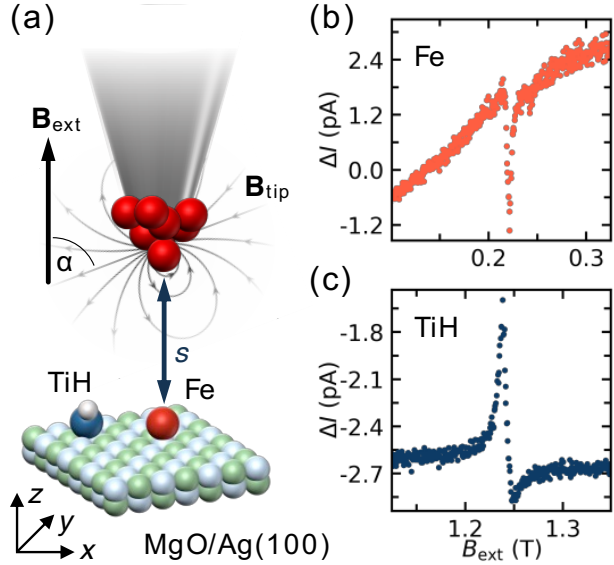

FIG. 1. Schematic of the experiment and typical EPR-STM spectra for Fe and TiH. (a) Single magnetic atoms of Fe and hydrogenated Ti (TiH) adsorbed on a bridge binding site on a double layer MgO on Ag (100) are subject to an external magnetic field $\mathbf{B}_{ext} \parallel z$ and a tip magnetic field $\mathbf{B}_{tip}$ from the magnetic STM tip. $\mathbf{B}_{ext}$ and the tip magnetic moment form an angle $\alpha$. The tip and surface magnetic moments are separated by a distance $s$. (b) Typical electron paramagnetic resonance spectrum of an Fe atom bound to an O atom at a constant radiofrequency of 36 GHz. (c) Same as (b) for TiH on a bridge binding site between two oxygen atoms (TiH).

The true magnetic moment

$$\mu = \frac{hf}{2\mathbf{e}_{spin} \cdot [\mathbf{B}_{ext} + \mathbf{B}_{tip}(s, \mathbf{B}_{ext})]}, \tag{3}$$

requires the precise knowledge of the magnetic field of the tip at the atom's position, $\mathbf{B}_{tip}(s, \mathbf{B}_{ext})$, where $s$ is the tip-sample distance and of the axis $\mathbf{e}_{spin}$ along which the magnetic states are eigenstates of the magnetic interaction term given by $\mathbf{B}_{tip}$, $\mathbf{B}_{ext}$, and the magnetic anisotropy. This axis can be defined by a strong uniaxial anisotropy term, as in the case of Fe, or by $\mathbf{B}_{ext}$ if the Zeeman energy is sufficiently strong compared to the local field of the tip, as is the case for TiH.

That is, a small tip magnetic field parallel to $\mathbf{e}_\text{spin}$ shifts the resonance line much more than the same field perpendicular to $\mathbf{e}_\text{spin}$. The difficulty in directly applying Eq. 3 to EPR-STM spectra arises from the fact that $\mathbf{B}_\text{tip}$ is hard to measure and control, because it depends on the structure and composition of the STM microtips [23]. Whereas magnetic stray fields from nearby magnetic assemblies on the surface can be minimized by choosing sufficiently isolated atoms on the surface, performing EPR-STM with magnetic tips has so far proven to be unavoidable to obtain reproducible EPR spectra, despite alternative proposals for using nonmagnetic tips [26].

In the past, three possible solutions to this problem have been presented:

(i) Circumvent the need to know $\mathbf{B}_\text{tip}$ by measuring the dipolar interaction between two magnetic surface atoms as a function of their distance and/or orientation of their connecting vector with respect to $\mathbf{B}_\text{ext}$ [16, 22]. The known distance-dependence of the dipolar interaction allows to directly obtain the magnetic moment if one moment is known or both are the same. This method is arguably the most accurate to date but requires either to manipulate atoms on the surface or to probe many different random dimer orientations and separations if the local properties of the atoms are the same. [29, 30]. Other complications arise if the electronic states of the two species hybridize when brought closely together [28]. Moreover, this approach requires detailed knowledge of the magnetic anisotropy of the EPR species to correctly estimate the dipolar interaction.

(ii) $\mathbf{B}_\text{tip}$ is characterized by acquiring a large EPR-STM data set that covers a wide range of tip-sample-distances and EPR conditions [23]. The full characterization of the tip vector field, however, is rather time consuming. An alternative is presented by a tip magnetization that is independent of the applied external magnetic field during the EPR sweeps, such as bulk magnetic tips [12] or paramagnetic tips that are typically saturated for large values of $B_\text{ext}$ (> 1T). In this case, the impact of the magnetic tip presents itself as a constant offset in EPR measurements at different radiofrequencies [10, 11].

(iii) Recording of EPR spectra at increasing tip-sample distance $s$ and extrapolation to infinite distance. This method suffers from two issues. Magnetic dipole interactions are long-range, which requires a precise model of how $\mathbf{B}_\text{tip}$ evolves with $s$ [22]. In addition, the EPR signal strongly reduces with $s$, rendering large distances not practical.

To circumvent the above difficulties, we propose to search for tip-sample distances, at which the tip magnetic field is minimized along $\mathbf{e}_\text{spin}$. $\mathbf{B}_\text{tip}$ is well described by the sum of a local exchange and a dipolar magnetic field. In the cartesian axes of the surface (see Fig. 1a) it is given by [13, 23],

$$\mathbf{B}_\text{tip} = \mathbf{B}_\text{xc} + \mathbf{B}_\text{dip} = \begin{pmatrix} B_{\text{xc},x}(s, \mathbf{B}_\text{ext}) + B_{\text{dip},x}(s, \mathbf{B}_\text{ext}) \\ B_{\text{xc},z}(s, \mathbf{B}_\text{ext}) + B_{\text{dip},z}(s, \mathbf{B}_\text{ext}) \end{pmatrix} = \rho(\mathbf{B}_\text{ext}) \begin{pmatrix} \left(ae^{-s/\lambda_\text{xc}} - \frac{b}{s^3}\right) \sin\alpha \\ \left(ae^{-s/\lambda_\text{xc}} + \frac{2b}{s^3}\right) \cos\alpha \end{pmatrix}$$

(4)

where the factor $\rho$ accounts for the relative changes of the tip polarization directly through $\mathbf{B}_\text{ext}$ as well as indirectly through a back-action of the surface atom on the tip state due to changes in the surface atom polarization with $\mathbf{B}_\text{ext}$. The coordinate system is oriented such that the y-component of the tip magnetization vanishes. Note that the dependence of $\rho$ on $\mathbf{B}_{\text{ext}}$ is the main complication for precisely determining atomic magnetic moments with EPR-STM. Further, Eq. 4 includes the sd-like exchange field strength $a$, the distance $s$ between the tip and surface magnetic moment, the exchange decay length $\lambda_{\text{xc}}$, and the angle $\alpha$ between the $z$-axis that is along the surface normal and the tip magnetization (Fig. 1a). The dipolar field strength $b = \mu_0 \mu_{\text{tip}}/4\pi$ is proportional to the vacuum permeability $\mu_0$ and the tip magnetic moment $\mu_{\text{tip}}$. Note that Eq. 4 assumes that the tip magnetic moment is located right above the surface magnetic moment, i.e., their connecting vector points along the out-of-plane direction $z$ (Fig. 1).

Thus, our task is to find the specific NOTIN tip-sample distances $s_0$ at which

$$\mathbf{B}_{\text{tip}}(s_0) \cdot \mathbf{e}_{\text{spin}} = 0. \quad (5)$$

From Eq. 4, we see that only trivial solutions exist for the strong condition $B_{\text{tip},x}(s_0) = B_{\text{tip},z}(s_0) = 0$, i.e., $|\mathbf{B}_{\text{tip}}| = 0$ for all $s$. This is because of the opposing signs of $B_{\text{dip},x}$ and $B_{\text{dip},z}$, whereas the components of $B_{\text{xc}}$ have the same sign. Thus, nontrivial solutions only exist if $\mathbf{e}_{\text{spin}}$ points either along $x$ or $z$. This situation occurs for surface atoms with strong uniaxial magnetic anisotropy, such as Fe on MgO, or for external magnetic fields $B_{\text{ext}} \gg B_{\text{tip}}$, both along $x$ or $z$, such as for TiH on MgO in our experiment. In these cases, our task reduces to finding solutions to either $B_{\text{tip},x}(s_{0,x}) = 0$ or $B_{\text{tip},z}(s_{0,z}) = 0$, respectively. Importantly, such solutions do exist for typical experimental conditions, as we will show below.

In the range of experimentally relevant parameters ($\lambda_{\text{xc}}, s > 0$), two nontrivial NOTIN distances exist depending on the relative sign of $a$ and $b$. If $\text{sign}(a) = \text{sign}(b)$, $B_{\text{tip},x}(s_0) = 0$ is fulfilled for the two values

$$s_{0,x}^{\pm} = -3\lambda_{\text{xc}} W_{0,-1}\left[-\frac{(b/a)^{1/3}}{3\lambda_{\text{xc}}}\right], \quad (6)$$

where $W_{0,-1}$ are the upper and the lower branch of the Lambert $W$ function [31], respectively, evaluated for the argument given in square brackets. On the other hand, if $\text{sign}(a) = -\text{sign}(b)$, $B_{\text{tip},z}(s_0) = 0$ has the two solutions

$$s_{0,z}^{\pm} = -3\lambda_{\text{xc}} W_{0,-1}\left[-\frac{(-2b/a)^{1/3}}{3\lambda_{\text{xc}}}\right]. \quad (7)$$

In addition, two more trivial yet experimentally very challenging solutions exist if $\alpha = 0°$ and $\mathbf{e}_{\text{spin}} \parallel \mathbf{y}$ or if $\alpha = 90°$ and $\mathbf{e}_{\text{spin}} \parallel \mathbf{z}$, in which case any value of $s$ solves Eq. 5. However, $\alpha$ is a parameter that is typically difficult to control in the experiment. In principle, the application of a vector magnetic field [18, 19, 32] would allow to access all four solutions in one experiment by forcing $\mathbf{e}_{\text{spin}}$ to point either along $x$ or $z$ if the anisotropy field can be exceeded. Note, however, that in general $s_0 = s_0(\mathbf{B}_{\text{ext}})$ since $\mathbf{B}_{\text{tip}}$ and $\mathbf{e}_{\text{spin}}$ might depend on $\mathbf{B}_{\text{ext}}$.

Experimentally, it is easier to measure the dependence of the resonance field on the tunneling junction conductance $G$ rather than the tip-sample distance $s$. The relation between $G$ and $s$ is well-described by $G(s) = G_0 \exp(-s/\lambda_G)$ where $G_0$ is the point-contact conductance and $\lambda_G$ the decay length of the tunnelling conductance [33]. Hence, we can rewrite Eq. 4 (4) as

$$\mathbf{B}_{\text{tip}}(\gamma) = \rho \begin{pmatrix} \left(a\gamma^m + \frac{n}{\ln(\gamma)^3}\right) \sin \alpha \\ \left(a\gamma^m - \frac{2n}{\ln(\gamma)^3}\right) \cos \alpha \end{pmatrix}, \tag{8}$$

where $\gamma = G/G_0$ is the normalized conductance, $m = \lambda_G/\lambda_{\text{xc}} > 0$, and $n = b/\lambda_G^3$. The value of $m$ is expected to be on the order of unity due to the common origin and of $\lambda_G$ and $\lambda_{\text{xc}}$ in the overlap of the orbitals of the adatom and tip. In the following, we will search for the NOTIN values of $G$ instead of $s$.

**RESULTS**

To test the proposed NOTIN-EPR method, we choose two well-studied yet magnetically distinct magnetic surface atoms as shown in Fig. 1, i.e., Fe and hydrogenated Ti atoms deposited on a bilayer of MgO grown on a Ag(100) surface. Fe is known to bind on top of an oxygen atom of the MgO and to have a strong out-of-plane magnetic anisotropy leading to a non-vanishing orbital moment $L = 2$ in addition to its spin moment $S = 2$ [20]. On the other hand, hydrogenated Ti on the bridge binding side (TiH) has a spin moment $S = 1/2$ and a negligible orbital moment along the out-of-plane direction [10, 22], but exhibits some variation of the magnetic moment in the plane [15, 18, 32]. Although Fe and TiH on MgO/Ag(100) are well studied, variations of their magnetic moment µ of up to 1 µ$_B$ for Fe can be found in the literature [16, 17, 19, 20]. The EPR-STM spectra were acquired by sweeping $B_{\text{ext}}$ at constant radiofrequency and recording the change in tunneling current that is caused by the radiofrequency signal generated by an antenna placed next to the STM tip. The radiofrequency signal was square-modulated at a frequency of 971 kHz and the changes in tunnelling current were measured using a lock-in technique [10]. To avoid stray fields from nearby magnetic surface atoms, we only study EPR species that are separated by more than 2 nm from other adsorbates. During EPR sweeps, the STM feedback loop is constantly engaged and an atom-tracking scheme stabilizes the tip on top of the adatom.

Figure 2 shows typical normalized EPR spectra recorded by sweeping the out-of-plane $\mathbf{B}_{\text{ext}}$ at different conductances $G$ and radiofrequencies. In each spectrum, a resonance can be identified that shifts with $G$. Further, the spectra show a nonlinear background $I_{\text{bg}}$ that depends on both $B_{\text{ext}}$ and $G$. The background of the spectra arises from rectified radiofrequency currents. Such a rectification occurs when the oscillation of the radiofrequency voltage around the set point voltage $V$ covers regions where $dG/dV \neq 0$ [10, 34]. Thus, a background signal that depends on $B_{\text{ext}}$ (see Figs. 1b and 2) implies $G = G(B_{\text{ext}})$, which is ascribed to a change of the magnetic state of the tip as well as of the surface atom when $B_{\text{ext}}$ is varied. The background changes are particularly pronounced for Fe due to the relatively low $B_{\text{ext}} \approx 100$ mT, which cannot saturate the magnetization of the paramagnetic tip. Thus, the magnetic background is expected to be described by the thermal occupation of the magnetic levels of the tip and adatom following a tanh function. Accordingly, the magnetic backgrounds in our EPR spectra are well described by a simple $\beta \tanh(pB_{\text{ext}})^2$ term with proportionality constants $\beta$ and $p$ that are independent fit parameters for each spectrum. It is important to note that the presence of magnetic-field dependent backgrounds in the EPR-STM spectra clearly demonstrates that the magnetic properties of the tunnel junction change with $B_{\text{ext}}$. This provides further motivation to use the NOTIN-EPR approach.

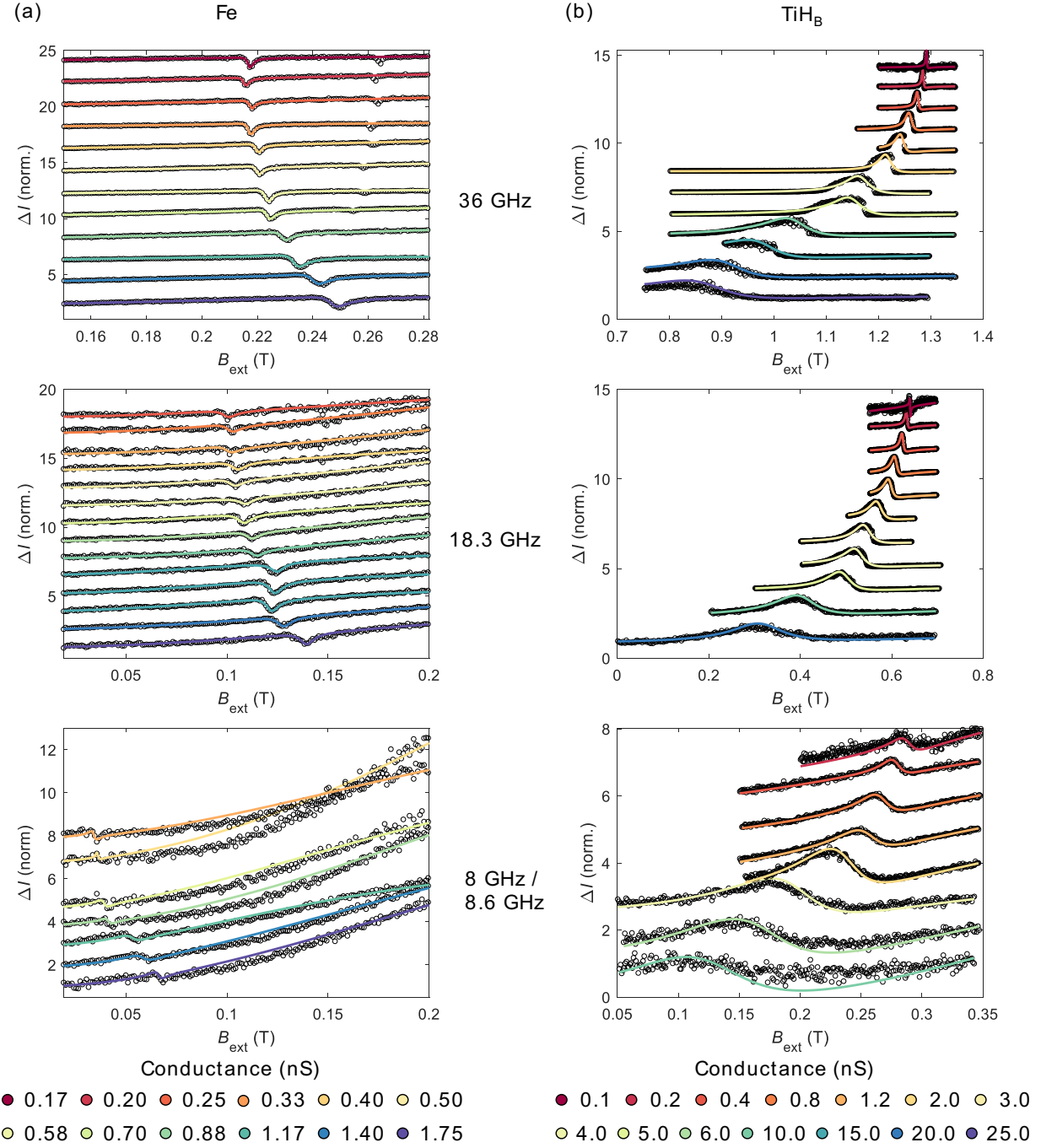

FIG. 2. EPR-STM spectra of (a) Fe and (b) TiH as a function of $G$ and $f$. The frequencies from top to bottom are 36, 18.3 and 8 GHz for Fe and 36, 18.3 and 8 GHz for TiH. The spectra are normalized to a peak-to-peak amplitude of 1 for better visibility. The resonance positions are obtained from fits using a Fano line shape (Eq. 9). All spectra are offset along the vertical axis for clarity.

To determine the external magnetic field at resonance $B_{\text{ext}}^0$, the EPR spectra are fitted with a Fano function [28]. The acquired current as given by the lock-in amplifier is

$$\Delta I = I_{\text{off}} + I_{\text{bg}}(B_{\text{ext}}, G) + I_{\text{EPR}} \frac{[q+(B_{\text{ext}}-B_{\text{ext}}^0)/\Gamma]^2}{1+[(B_{\text{ext}}-B_{\text{ext}}^0)/\Gamma]^2} \tag{9}$$

where $I_{\text{off}}$ is a constant offset, $I_{\text{bg}}$ is the background depending on the magnetic field, $I_{\text{EPR}}$ is the peak amplitude, $q$ is the asymmetry parameter, and $\Gamma$ denotes the width of the peak. Note that a direct link between the magnetic-field-dependent background $I_{\text{bg}}$ and the polarization factor $\rho$ in Eq. 4 cannot be made since it requires to disentangle the change of the tip and the surface atom polarization with $B_{\text{ext}}$, which would increase the number of fit parameters considerably. Further, for Fe measured at 8 GHz, we excluded 6 out of 14 spectra from the analysis because the resonance could not be resolved within our signal to noise ratio.

The curves of $\tilde{\mu}$ (extracted by Eq. 2) as a function of $G$ for different $f$ are plotted in Fig. 3. Two distinct behaviours are immediately apparent: For Fe, the three curves cross at $G \approx 1.1$ nS, whereas the data for TiH suggest an intersection point just outside the measurement window close to $G = 0$. Further, the curves are not linear in $G$ as can be noted also from Eq. 8, which is due to the dipolar contribution to the tip magnetic field as well as to the parameter $\rho$. Considering this nonlinear behavior, we fit the $B_{\text{ext}}^0(G)$ curves with a third-order polynomial function, which is sufficient in most cases. This allows us to determine the precise crossing point and, thus, the NOTIN $\tilde{\mu}$. Accordingly, for Fe we extract a magnetic moment at the intersection point of $\mu_{\text{Fe}} = 5.47 \pm 0.06$ $\mu_B$ along the $z$-direction. The error is obtained from the confidence interval of one standard deviation $\sigma$ derived from the polynomial fits (see shaded areas in Fig. 3a) and by computing the resulting error as $\sqrt{\sum_{i=1}^{n} \sigma_i^2}/n$ at the intersection point, where $n = 3$ for the three different radiofrequencies. For TiH, on the other hand, an extrapolation is required to obtain $\mu_{\text{TiH}}$. To this end, we fit the apparent moment $\tilde{\mu}$ for TiH also with a third-order polynomial and obtain an intersection point close to $G = 0$, which means that only at a large tip-sample distance the condition $B_{\text{tip}} = 0$ is fulfilled for this specific microtip. We thus find $\mu_{\text{TiH}} = 0.99 \pm 0.01$ $\mu_B$.

Note that a fit based on our full model (Eq. 8) would introduce more fitting parameters, and therefore, increase the uncertainty in the fit. An even more advanced analysis including a changing $\mathbf{e}_{\text{spin}}$ for the surface atom with tip-sample distance and $\mathbf{B}_{\text{ext}}$ would further introduce $\alpha$ as a free parameter in Eq. 8 and, thus, require more data for a unique fit. From the data in Fig. 3, a single crossing point can be extracted for each atom. In principle, a second crossing point exists [Eqs. 6 and 7] but the specific tip magnetic properties and the experimental stability of the tip-atom tunnel junction determines if both crossing points can be detected.

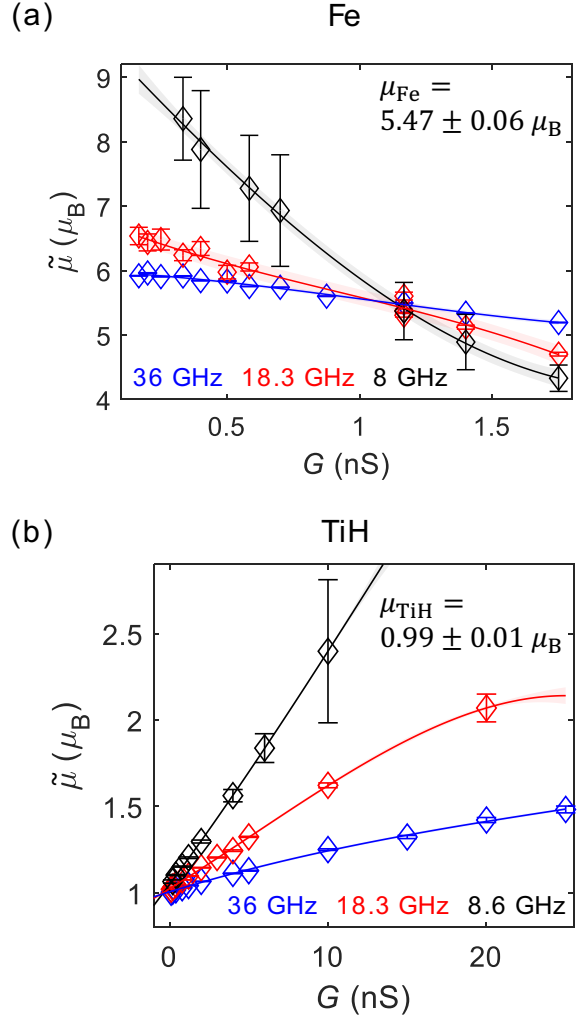

FIG. 3. Apparent magnetic moment as a function of the junction conductance measured for (a) Fe and (b) TiH at different radiofrequencies. Solid lines are fits by a third-order polynomial function with the corresponding uncertainty given by the shaded areas. The error bars are obtained from the Fano-lineshape fits to the data in Fig. 2 and accounted for during the polynomial fits. The true magnetic moment of the adatom species can be obtained at the intersection point of the curves.

## DISCUSSION

**Comparison to previous measurements of µ**

We can compare our results for the magnetic moment here with previous reports of the same systems. For Fe atoms, an x-ray magnetic circular dichroism experiment revealed a magnetic moment of 5.2 $\mu_B$ [20] though containing uncertainties from the analysis tools as well as averaging over many Fe atoms and possibly also multimers on 5-6 monolayers of relatively rough MgO. An EPR-STM study deduced $\mu = 5.44 \pm 0.03\ \mu_B$ from the distance dependence of the dipolar interaction between two Fe atoms [16]. Another EPR-STM study reported $\mu = 5.23 \pm 0.47\ \mu_B$ by measuring the temperature evolution of relative intensities between two dimer peaks [17]. Finally, Willke and coworkers extracted $\mu = 5.35 \pm 0.14$ $\mu_B$ and $\mu = 4.29 \pm 0.79$ $\mu_B$ by performing frequency sweeps and tip-height sweeps without external magnetic field using EPR-STM [19]. Thus, the reported values for the magnetic moment of Fe measured by EPR-STM differ by more than 1 $\mu_B$ although each individual measurement claims a smaller error. The methodologically accurate distance-dependent measurement of Fe dimers [16] agree well with our results of $\mu_{Fe} = 5.47 \pm 0.06$ $\mu_B$.

For TiH, we focus on reports of the out-of-plane magnetic moment because we apply $B_{ext}$ along this direction. A much smaller spread of its magnetic moment is reported. Natterer and coworkers found $\mu = 1.004 \pm 0.001$ $\mu_B$ [11], consistent with another study that reported $\mu = 1.00 \pm 0.01$ $\mu_B$ [10]. Both works performed EPR-STM at $B_{ext} \geq 1$ T, which makes the extraction procedure of $\mu$ more reliable (see method (ii) above) because the tip magnetization induces, to a good approximation, a constant offset magnetic field. Using a vector magnetic field, Kim et al. [18] reported $\mu = 0.99 \pm 0.01$ $\mu_B$ for the out-of-plane direction. All values are in line with our measurement.

We note that our accuracy for the magnetic moments is about $10^{-2}$, which is comparable to the most accurate results obtained up to now by EPR-STM. However, our measurements remove one of the major sources of systematic errors, namely the tip magnetic field. Other sources of systematic errors might include mechanical instabilities, inaccurately calibrated magnetic field sensors or trapped fluxes in the superconducting coils. The statistical error can, in principle, be reduced by increasing the measurement time. We note also that reaching an accuracy comparable to that of the most precise EPR measurements is challenging in an STM, because tip vibrations of the order of 1 pm lead to a broadening of the EPR spectra in the mT range [23]. Under these conditions, EPR-STM is limited to an accuracy of about $10^{-3}$ unless the NOTIN approach is brought to the next level of accuracy by requiring Eq. 5 and $d\mathbf{B}_{tip}/ds |_{s=s_0} \cdot \mathbf{e}_{spin} = 0$ to be simultaneously fulfilled. This stricter condition must not prevent the EPR driving as in the case of Fe, where field gradients along $\mathbf{e}_{spin}$ induce the EPR transitions [23]. However, for half-integer spin systems such as TiH, the stricter condition could be fulfilled within our model (Eq. 4) if the NOTIN distances $s_{0,x/z}$ given by Eqs. 6 and 7 are simultaneously equal to $\lambda_{xc}/3$, which is, however, challenging given that $\lambda_{xc}$ is typically of the order of 100 pm [23].

**Requirements to apply NOTIN-EPR**

Finding the NOTIN tip-sample distances, at which the impact of the tip magnetic field on the resonance field is minimized is in principle possible if the following conditions are fulfilled: For a surface atom with a strong uniaxial magnetic anisotropy such as Fe [20], (I) dipolar and exchange contributions to $\mathbf{B}_{tip}$ either oppose each other (Eq. 4) or $\mathbf{B}_{tip}$ points perpendicular to $\mathbf{e}_{spin}$ at certain tip sample distances (Eq. 5). For a surface atom with a weak magnetic anisotropy compared to $B_{tip}$ and $B_{ext}$ such as TiH, (I) must apply and (II) the tip magnetic field must have a minor impact on $\mathbf{e}_{spin}$, i.e., $B_{ext} \gg B_{tip}$. In any case, EPR-STM must be feasible at the NOTIN tip-sample distance $s_0$, which requires a sufficiently strong EPR signal even at large $s_0$ or a sufficient stability of the surface atom for a small $s_0$. Depending on the specific experimental conditions, one of the two possible values of $s_0$ (Eqs. 6 or 7) might be accessible. As our results demonstrate, conditions (I) and (II) can be fulfilled in the experiment, which allows one to extract magnetic moments with a high precision by circumventing a major source of systematic errors, the tip magnetic field. The NOTIN-EPR method is expected to be more precise if different radiofrequencies are chosen such that the $\tilde{\mu}(G)$ curves (Fig. 3) intersect with significantly different slopes. We note that if the NOTIN distance $s_0$ cannot be reached, a model-based extrapolation (Eqs. 3 and 4) is required to obtain the magnetic moment. Using a vector magnetic field [12, 18, 19] can further simplify the condition of Eq. 3 because it would allow one to minimize the impact of $\mathbf{B}_{tip}$ on the resonance condition by aligning $\mathbf{e}_{spin}$ via $\mathbf{B}_{ext}$ perpendicular to $\mathbf{B}_{tip}$.

**Bistable magnetic tips**

Finally, we demonstrate an alternative method to cancel the influence of $\mathbf{B}_{tip}$ on the magnetic moments extracted from the EPR spectra. This method confirms the NOTIN results and provides another simple tool to measure the magnetic moment of surface atoms using EPR-STM. Careful inspection of the data for Fe at 36 GHz in Fig. 2a reveals a second resonance for $B_{ext}$ larger than the main resonance field at about 260 mT. We assign this double peak resonance to a bistable magnetic tip, which switches between two metastable states on time scales much faster than the measurement time of EPR spectra, typically a few minutes in our experiments. Similar features were also observed recently by other groups (see Supplementary Information of Ref. [13] and Ref. [24]). Indeed, a more thorough analysis of all EPR-active microtips used in our experiments over the last two years (about 1000 microtips), reveals that about 1 % of EPR-active tips are magnetically bistable. Figure 4 shows examples of EPR spectra recorded with bistable tips. By taking the average of the two resonance fields, the effects of $\pm\mathbf{B}_{tip}$ compensate each other, resulting in an intrinsic resonance field consistent with the magnetic moment obtained by the NOTIN method. Using in total 7 bistable tips for Fe and 19 for TiH, we find magnetic moments of $\mu_{Fe} = 5.4 \pm 0.1$ and $\mu_{TiH} = 0.98 \pm 0.03$, respectively. We note that each EPR sweep with a bistable magnetic tip is susceptible to statistical and systematic errors such as mechanical instabilities or trapped fluxes in the superconducting coils that might result in deviations from the nominal magnetic field of about 1 mT as observed in Fig. 4b (upper panel). The respective uncertainties in magnetic moments could be further minimized by increasing the measurement time or the number of measured bistable tips. The good agreement of the extracted magnetic moments compared with the NOTIN approach implies that the magnetization of the tip switches between two opposite orientations, indicating that all these bistable magnetic tips have a uniaxial magnetic anisotropy. To favour a bistability of the magnetic tip, low external magnetic fields and correspondingly low radiofrequencies are beneficial. We note, however, that we have currently not found a reliable procedure to obtain bistable magnetic tips. To summarize, the good agreement between these two methods to extract the magnetic moment of a surface atom further underlines their validity.

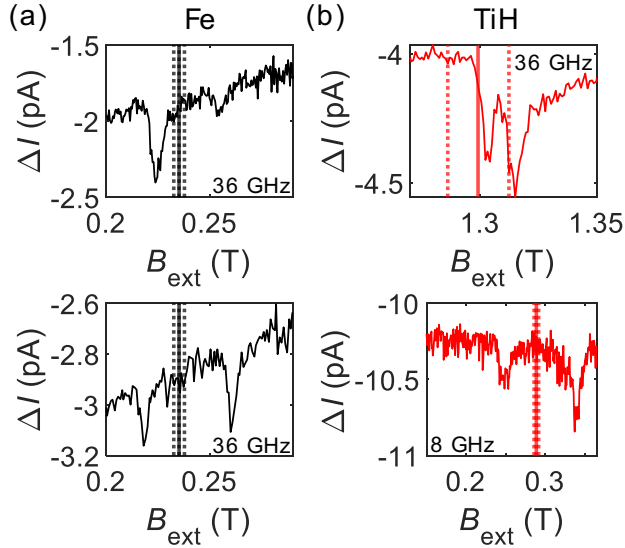

FIG. 4. EPR-STM spectra obtained with bistable magnetic tips on Fe (a) and on TiH (b) at the indicated radiofrequencies. Solid vertical lines indicate the expected intrinsic resonance fields based on the magnetic moments extracted from Fig. 3. Dashed vertical lines mark the respective uncertainty.

## CONCLUSIONS

We introduced two methods to measure the magnetic moment of a surface atom with increased precision using EPR-STM based on NOTIN-EPR and bistable magnetic tips. Both methods yield consistent results with the literature, but offer higher precision by minimizing tip-related systematic errors. In addition, these methods can be applied to a large class of EPR species, irrespective of the magnitude of the magnetic moments or the size of the probed species, such as magnetic molecules and magnetic nanostructures.

Further, we showed that magnetic field dependent backgrounds in the EPR spectra, which are especially pronounced at low values $B_{ext}$, can be well described by a $\tanh(pB_{ext})^2$ function with a proportionality constant $p$. This result will prove especially useful in the study of high spin species such as 4f elements [35] that typically require low external magnetic fields. Ultimately, by using a well-known surface magnetic species, the magnetic background might yield valuable insights on the magnetic microstructure of the STM tip.

Future studies might aim at implementing a feedback scheme based on radiofrequency modulation and a sweeping of $B_{ext}$ to automatically find the NOTIN tip-sample distance, at which the extracted magnetic moment is independent of the radiofrequency. Also, investigations on the dependence of the NOTIN points on the details of the magnetic tips present an interesting topic for future works. In general, NOTIN-EPR is not limited to magnetic tips that show dipolar and exchange fields but is expected to work as well for tips featuring different exchange regimes [36]. The NOTIN technique could also be useful for spin-polarized STM studies in general, where the impact of the tip magnetic field on the object of interest on the surface is typically uncharacterized.

## ACKNOWLEDGMENTS

We acknowledge funding from the Swiss National Science Foundation, Project No. 200021_163225.